\documentclass{article}
\usepackage{spconf,amsmath,graphicx}
\usepackage{multirow}
\usepackage{makecell}
\usepackage[flushleft]{threeparttable}
\usepackage{hyperref}
\usepackage{amsmath}
\usepackage{soul,color}

\usepackage{enumitem}
\usepackage{xcolor}
\usepackage{amssymb}
\usepackage{pifont}
\usepackage{booktabs}

\title{Mixer-TTS: Non-Autoregressive, Fast and Compact Text-To-Speech Model conditioned on Language Model embeddings}

\name{Oktai Tatanov, Stanislav Beliaev, Boris Ginsburg
\thanks{Preprint. Submitted to ICASSP-22.}
}
\address{NVIDIA, Santa Clara}

\begin{document}

\maketitle

\begin{abstract}
This paper describes Mixer-TTS, a non-autoregressive model for mel-spectrogram generation. The model is based on the MLP-Mixer architecture adapted for speech synthesis. The basic Mixer-TTS contains pitch and duration predictors, with the latter being trained with an unsupervised TTS alignment framework. Alongside the basic model, we propose the extended version which additionally uses token embeddings from a pre-trained language model. Basic Mixer-TTS and its extended version achieve a mean opinion score (MOS) of 4.05 and 4.11, respectively, compared to a MOS of 4.27 of original LJSpeech samples. Both versions have a small number of parameters and enable much faster speech synthesis compared to the models with similar quality.

\end{abstract}


\section{Introduction}
\label{sec:intro}

Recent neural text-to-speech (TTS) models have significantly improved the speed, robustness, and quality of generated speech. The improvement in training and inference speed is mostly related to switching from sequential, autoregressive models~\cite{Tacotron1, Tacotron2, DeepVoice3}) to parallel, non-autoregressive models~\cite{Fastspeech2019,fastspeech2,2021fastpitch, beliaev2021talknet}. Non-autoregressive models can generate speech two orders of magnitude faster than auto-regressive models with similar quality. For example, FastPitch generates mel-spectrograms 60x faster than Tacotron 2~\cite{2021fastpitch}. 

The robustness of TTS models was significantly improved by using an explicit duration predictor~\cite{DeepVoice1,DeepVoice2,Fastspeech2019,fastspeech2,2021fastpitch,beliaev2021talknet} that practically eliminates skipping and repeating words, which were common issues in popular models like Tacotron 2. Traditionally, models with duration predictors have been trained in a supervised manner with external ground truth alignments. For example, TalkNet used the alignment from auxiliary ASR models, while FastSpeech and FastPitch used alignments from a teacher TTS model. Glow-TTS~\cite{kim2020glowtts}  proposed a flow-based algorithm for unsupervised alignment training. This algorithm has been improved in RAD-TTS~\cite{radtts2021} and modified for non-autoregressive models in~\cite{badlani2021one}. This new alignment framework greatly simplifies TTS training pipeline.

% Prosody 
% 1. pitch
The quality of speech generated by the first non-autoregressive models was inferior to state-of-the-art autoregressive models. FastPitch~\cite{2021fastpitch}, a non-autoregressive model, closed the gap in quality by a adding pitch predictor for fundamental frequency (F0).
% 2. LM embeddings
Hayashi et al~\cite{hayashi19_interspeech} proposed to augment TTS model with input representation from a pre-trained BERT~\cite{bert2019} language model (LM). The authors hypothesized that text embeddings contain information about the importance of each word, which helped to improve speech prosody and pronunciation. The usage of semantic context for TTS was extended in~\cite{xu2021improving}.

% Model overview 
% 0. improve quality 
% 1. decrease model size
% 2. increase inference speed
In this paper, we present Mixer-TTS, a non-autoregressive model for text to mel-spectrogram synthesis. The model backbone is based on the MLP-Mixer~\cite{tolstikhin2021mlp} architecture from computer vision  adapted for speech. The new backbone makes the model significantly smaller and faster than Transformer-based TTS models \cite{fastspeech2, 2021fastpitch}.
Our model uses an explicit duration predictor, which is trained by the unsupervised alignment framework proposed in \cite{badlani2021one}. Mixer-TTS combines two methods to improve the prosody of generated speech. The basic version has an explicit pitch predictor similar to FastPitch~\cite{2021fastpitch}. The extended version adds token embeddings from an external pre-trained LM to improve speech prosody and pronunciation. Using token embeddings is significantly less expensive than inferring BERT outputs as in~\cite{hayashi19_interspeech}. They notably improve speech quality with a very modest increase in the model size and inference speed. 

We evaluate the quality of the proposed models combined with a HiFi-GAN~\cite{kong2020hifigan} vocoder on LJSpeech~\cite{ljspeech17}. Mixer-TTS achieves a mean opinion score (MOS) of 4.05 compared to a MOS of 4.27 for the original speech samples. The extended model with LM embeddings improves MOS to 4.11. The basic version has 19.2M parameters, and the extended model has 24M. Mixer-TTS samples are published online\footnote{\url{https://mixer-tts.github.io/}}.

\section{Model Architecture}
\label{sec:proposedmodel}

\begin{figure}[tb]
  % \centering
  \includegraphics[width=1.00\linewidth]{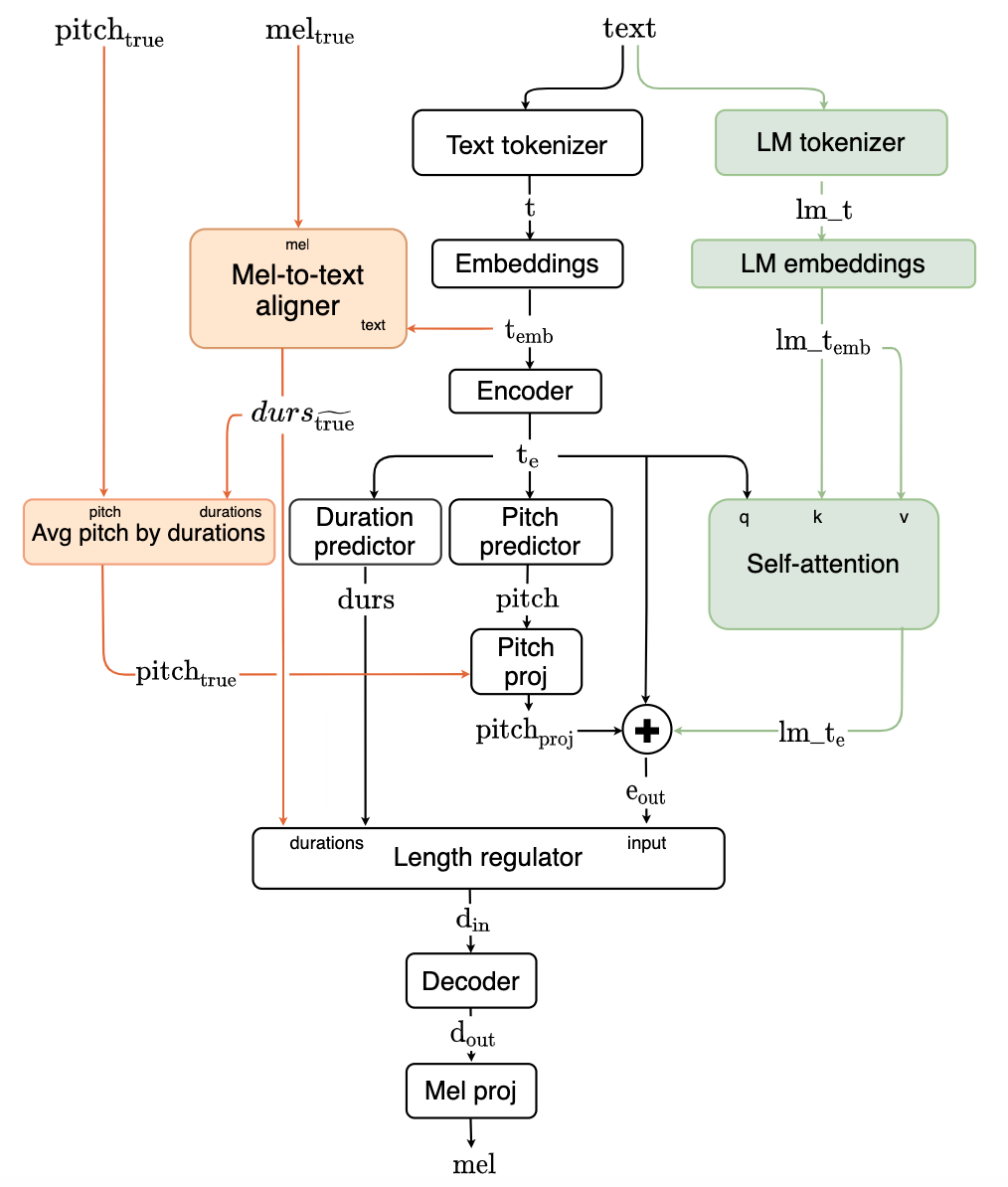}
  \caption{Training and inference pipeline of Mixer-TTS. During training, the decoder uses durations from mel-to-text aligner and ground truth pitch. During inference, durations and pitch are obtained from predictors. Extended Mixer-TTS contains additional blocks for LM embeddings.}
  \label{fig:model}
\end{figure}

The model architecture is shown in Figure~\ref{fig:model}. We encode the text and align it by using audio features in a separate module to get “ground truth” durations. Then, we calculate character or phoneme-level pitch values and feed them all into the length regulator module to expand each character or phoneme feature along with their corresponding durations. Next, the decoder generates mel-spectrogram from the encoded representations.

The basic Mixer-TTS is structurally similar to FastPitch with two major changes. First, we replaced all feed-forward Transformer-based blocks in the encoder and decoder with new Mixer-TTS blocks (see subsection~\ref{ssec:mixerttsblock}). Second, we used an unsupervised speech-to-text alignment framework to train the duration predictor (see subsection~\ref{ssec:meltotextalignment}). 
The extended Mixer-TTS additionally includes conditioning on embeddings from pretrained LM (see subsection~\ref{ssec:nlpaligner}). We use the duration and pitch predictor architectures described in FastPitch.

The model is trained with loss function combined from aligner loss and mean-squared errors between ground-truth and predicted values for mel-spectrogram, duration and pitch:
\begin{align*}
& L = L_{aligner} + L_{mel} + 0.1 \cdot L_{durs} + 0.1 \cdot L_{pitch}
\end{align*}

\subsection{Mixer-TTS Block}
\label{ssec:mixerttsblock}

\begin{figure}[tb]
  \centering
  \includegraphics[width=0.76\linewidth]{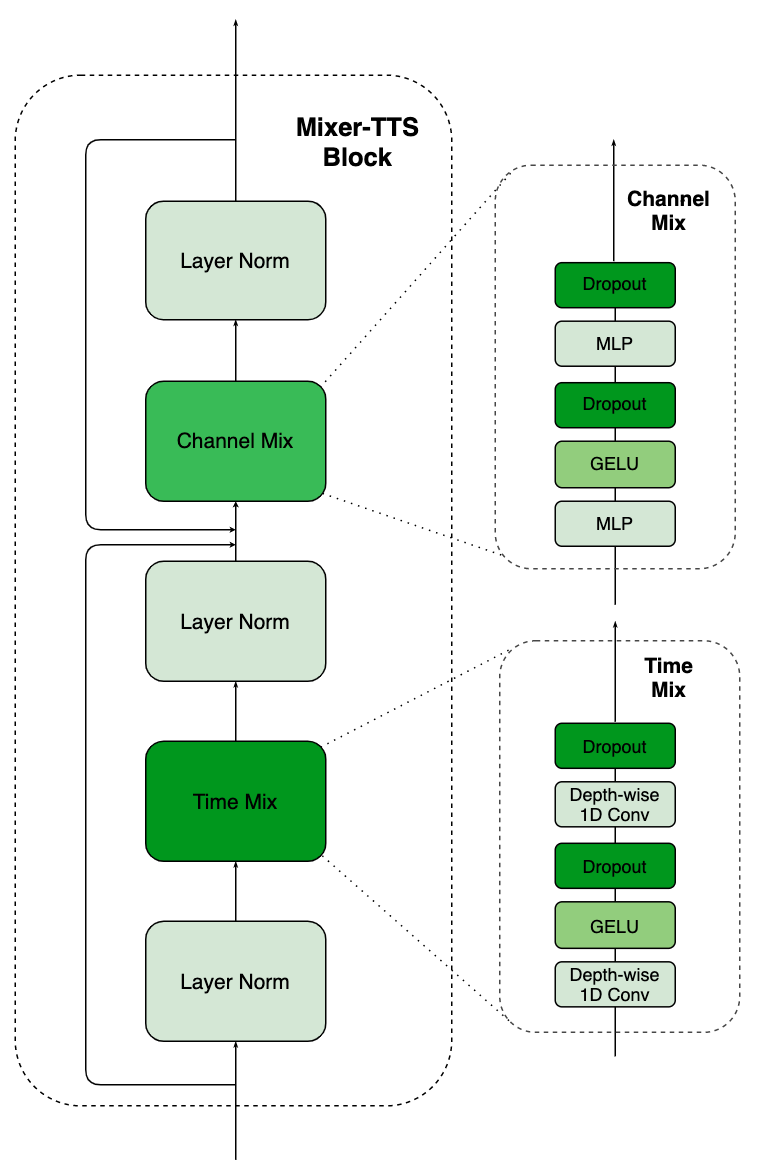}
  \caption{Mixer-TTS block consists of time and channel mix blocks. Channel-mix block includes two MLPs and a GELU. Time-mix block is similar but with depth-wise 1D convolutions instead of MLPs.
  }
  \label{fig:only_mixs}
\end{figure}

The MLP-Mixer architecture, introduced in \cite{tolstikhin2021mlp} for computer vision, is based exclusively on multi-layer perceptrons (MLPs). MLP-Mixer performs two key actions over input: “mixing” the per-location features and “mixing” spatial information. Both operations are implemented by the stack of two MLPs layers. The first MLP layer increases number of channels by an "expansion factor", and the second MLP layer reduces channels to the original value. However, such an approach is only possible when the input size for a layer is fixed by every dimension. To use this architecture for text-to-speech (i.e. case where one of the input's dimensions has dynamic size), we use "time-mixing", replacing MLPs with depth-wise 1D convolutions\footnote{In this case, model consists not only of MLPs, so we decided to call our block as \textsl{Mixer-TTS}, because the idea of "mixing" still remained.} and borrowed the original layer for channel "mixing". The rest of the structure of the original MLP-Mixer remains unchanged, including layer normalization and residual connections (see Figure~\ref{fig:only_mixs}). During mini-batch training and inference, when sequences in a batch are padded to match the longest sequence, we use sequence masking after MLP and depth-wise 1D convolution layers.

The encoder is composed of six stacked Mixer-TTS blocks with convolution kernel in time-mix growing linearly from $11$ to $21$ with step $2$. The decoder is composed of nine stacked Mixer-TTS blocks with kernel sizes growing from $15$ to $31$ in the same manner. Feature dimension is constant $384$ for all blocks used, channel-mix expansion factor is $4$ and there is no expansion factor in time mix. We used a dropout of $0.15$ in each block.

\subsection{Speech-to-text alignment framework}
\label{ssec:meltotextalignment}

Most non-autoregressive TTS models with duration prediction rely on durations extracted from external sources. However, in our work, we train the speech-to-text alignments jointly with the decoder by using adaptation of unsupervised alignment algorithm~\cite{badlani2021one} which was proposed in implementation of FastPitch 1.1\footnote{\url{https://github.com/NVIDIA/DeepLearningExamples/tree/master/PyTorch/SpeechSynthesis/FastPitch}}. This aligner encodes text and mel-spectrogram using 1D convolutions and projects them to a space with the same dimensionality. The "soft" alignment is computed using a pair-wise $L_2$ distance of the encoded text and mel representations, then Connectionist Temporal Classification (CTC) loss is used to learn these alignments. To obtain monotonic binarized alignments (i.e “ground-truth” durations), the Viterbi algorithm is used to find the most likely monotonic path. More details about the alignment framework can be found in \cite{badlani2021one}.

\subsection{Extended Mixer-TTS}
\label{ssec:nlpaligner}

The extended model takes advantage of token embeddings from external LM. We used ALBERT~\cite{lan2020albert} model from HuggingFace~\cite{wolf-etal-2020-transformers} pretrained on large corpus of English text. We kept the LM tokenization method and utilized frozen embeddings for input tokens.
The lengths of original and tokenized text from external LM are different because they are produced by different tokenizers. To align two sequences, we use a single head self-attention block applied to LM embeddings $lm_{emb}$ and encoder output $t_e$, which mixes their features while preserving the lengths of "basic" text embeddings.
Text features for self-attention aligning are extracted with convolutional layers preceded by separate positional embedding layers.

\section{Results}
\label{sec:results}

\subsection{Training details}
\label{ssec:training}

The model was trained on the LJSpeech dataset which was split into three sets: $12,500$ samples for training, $100$ samples for validation, and $500$ samples for testing. The text was lower-cased while leaving all punctuation intact. 
We experimented with two tokenization approaches: character-based and phoneme-based. For grapheme-to-phoneme conversion we used the ARPABET representation in the \textit{CMUdict}\footnote{\url{http://www.speech.cs.cmu.edu/cgi-bin/cmudict}} vocabulary and left ambiguous words and heteronyms in character representation.
We converted ground truth 22050Hz sampling rate audios to mel-spectrograms using a Short-Time Fourier Transform (STFT) with $50$ ms Hann window and $12.5$ ms frame hop. Ground truth pitch was extracted using the \textit{librosa} library~\cite{mcfee2015librosa} with values-aligned along mel-spectrogram frames.

The model was trained for $1000$ epochs using the LAMB optimizer~\cite{you2019large} with $\beta_1=0.9,\beta_2=0.98,\epsilon=10^{-8}$, a weight decay of ${10}^{-6}$ and gradient clipping of $1000.0$. A Noam annealing learning rate policy was used with a learning rate of $0.1$ and a $1000$ steps warmup. We used a total batch of $128$ for four GPUs with gradients accumulation of $2$. The training takes around $12$ hours on four V100 GPUs in mixed precision mode~\cite{micikevicius2017mixed}.

\subsection{Speech quality evaluation}
\label{ssec:evaluation}

We have conducted several mean opinion score (MOS) studies for generated speech quality comparison using Amazon Mechanical Turk\footnote{\url{https://www.mturk.com}}.
For evaluation, we selected Mturk workers with top performance ($\geq 95\%$ HITS Approval, $\geq 5000$ HITS Total), from the US only and with minimum high school degree.
We tested $50$ audio samples per model with $15$ people per sample. The scores ranged from $1.0$ to $5.0$ with a step of $0.5$. 

First, we compared Mixer-TTS with different tokenization approaches in combination with LM embeddings conditioning. The results are in Table~\ref{tab:mos-nlp}. Phonetic input representation slightly outperforms characters for the basic model. But for the extended model, the combination of character-based tokenization with LM embeddings performs better.

\begin{table}[tb]
\centering

\scalebox{1.00}{
    \begin{tabular}{c c r} 
    \toprule
    \textbf{LM embeddings} &
    \textbf{Tokenizer} &
    \textbf{MOS} \\
    \midrule
    \ding{51} & chars & $\mathbf{4.11 \pm 0.06}$ \\
    \ding{51} & phonemes & $4.06 \pm 0.06$ \\
    \ding{55} & phonemes & $4.06 \pm 0.06$ \\
    \ding{55} & chars & $4.03 \pm 0.06$ \\
    \bottomrule
    \end{tabular}
}
\caption{\textbf{Mixer-TTS Ablation Studies} with $95\%$ confidence interval. Four options of Mixer-TTS were used: with or without LM embeddings and with char-based or phoneme-based tokenizers.}
\label{tab:mos-nlp}
\end{table}

\begin{table}[ht]
\centering
\scalebox{1.00}{
    \begin{tabular}{l c r} 
    \toprule
    \textbf{Model} &
    \textbf{MOS} \\
    \midrule
    Ground truth audios & $4.27 \pm 0.05$ \\
    Ground truth mels & $4.18 \pm 0.07$ \\
    \midrule
    FastPitch & $4.05 \pm 0.06$ \\
    Tacotron 2 & $3.95 \pm 0.06$ \\
    TalkNet 2 & $3.95 \pm 0.07$ \\
    \midrule
    Mixer-TTS-X & $\mathbf{4.11 \pm 0.06}$ \\
    Mixer-TTS & $\mathbf{4.06 \pm 0.06}$ \\
    \bottomrule
    \end{tabular}
}
\caption{\textbf{Mean Opinion Scores (MOS)} with $95\%$ confidence interval. We used the HiFi-GAN vocoder which was trained on ground-truth mel-spectrograms and additionally fine-tuned for $100$k steps on outputs from every model. TalkNet 2 and FastPitch were trained with an unsupervised aligner framework to match the Mixer-TTS approach.}
\label{tab:mos-cmp}
\end{table}

As the main study, we compare Mixer-TTS with the following popular and relevant models: Tacotron 2, FastPitch, and TalkNet 2. The last two models were trained with the same aligner mechanism instead of the usual external set of durations to match the approach presented in Mixer-TTS. According to the results, the basic version of Mixer-TTS with phonemes achieves a comparable to FastPitch level of speech quality, and the extended version (Mixer-TTS-X) exceeds the quality of all the examined models (see Table~\ref{tab:mos-cmp}).

\subsection{Inference performance}
\label{ssec:speed}

% Environment Description
We compared Mixer-TTS inference with FastPitch as the fastest examined model. The measurement was done with a variable-length text generator based on snippets from the LJSpeech test set and batch size of one. Evaluation was done using AMP (Automatic Mixed Precision) in PyTorch $1.11$ with CUDA $11.3$, cuDNN $8.2$ and NVIDIA's A100 GPU. We measured the wall-time of mel-spectrogram generation starting from the raw text processing step and averaged results over $10$ consecutive runs with a warmup for cuDNN to adjust algorithms to input sizes.

\begin{figure}[t]
   \centering
  \includegraphics[width=0.95\linewidth]{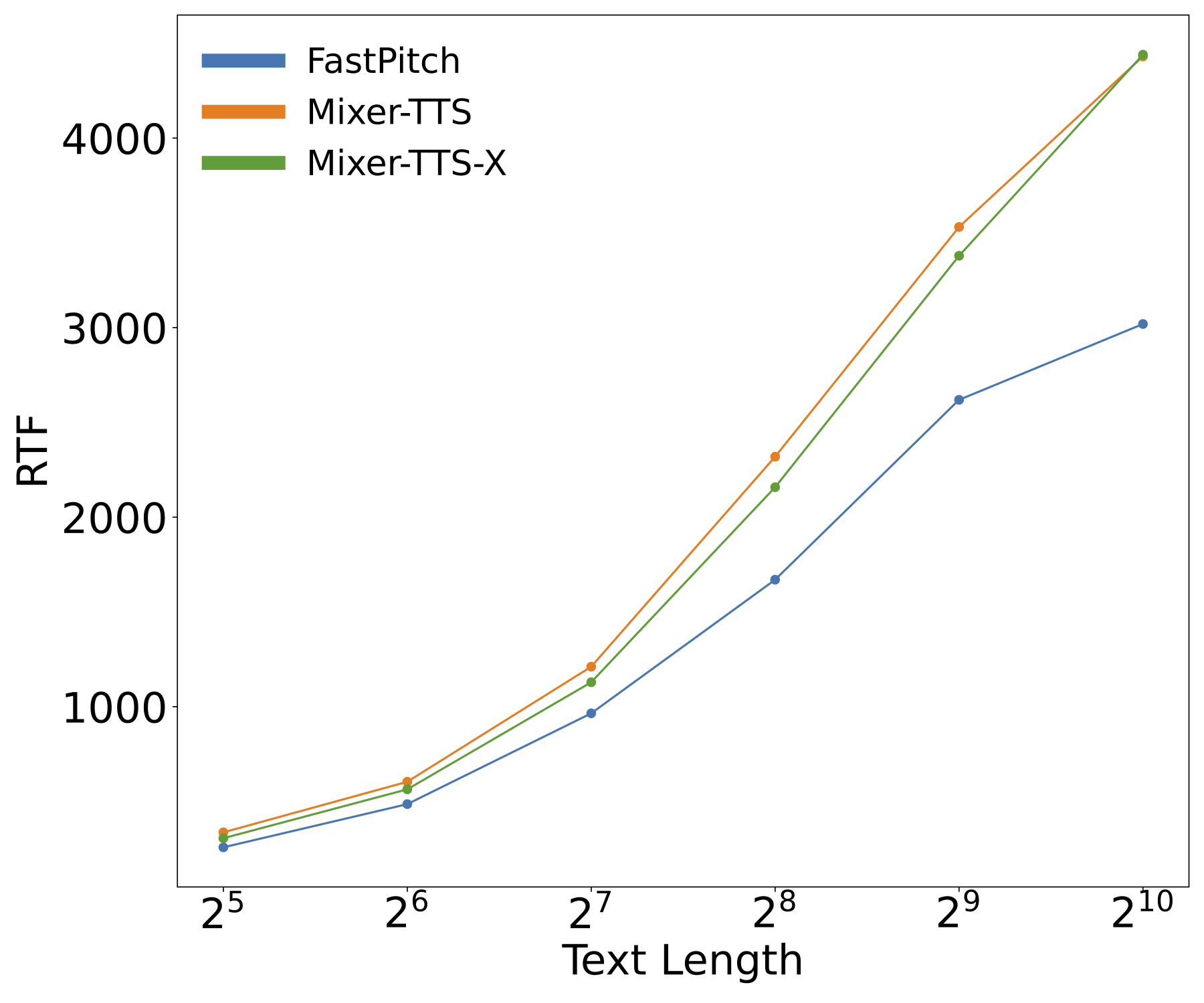}
  \caption{\textbf{Real-time factor (RTF) for mel-spectrogram generation.} We used a batch size equal to one with mixed precision on A100 GPU.}
  \label{fig:rtf}
\end{figure}

% Results Description
Mixer-TTS inference is notably faster than FastPitch and it scales better with increasing the input length (see Figure~\ref{fig:rtf}). Furthermore, the best version of our model has only 24 million parameters while FastPitch has 45 million parameters.

% \begin{table}[t]
% \caption{\textbf{Real-time factor (RTF) for mel-spectrogram generation.} We used a batch size equal to one with FP16 precision on A100 GPU.}
% \label{tab:rtf}
% \centering
% \scalebox{0.85}{
%     \begin{tabular}{r r r r }
%     \toprule
%     % \textbf{GPU} &
%     \textbf{\thead{Text length,\\char}} & 
%     \textbf{\thead{FastPitch,\\RTF}} &
%     \textbf{\thead{Mixer-TTS,\\RTF}} &
%     \textbf{\thead{Mixer-TTS-X,\\RTF}} \\
%     \midrule
%      32  & $258$  & $337$ / $1.30$x & $306$ / $1.19$x \\
%      64  & $486$  & $604$ / $1.24$x & $564$ / $1.17$x \\
%      128  & $965$  & $1211$ / $1.26$x & $1129$ / $1.17$x \\
%      256  & $1670$  & $2319$ / $1.39$x & $2158$ / $1.29$x \\
%      512  & $2619$  & $3531$ / $1.35$x & $3379$ / $1.29$x \\
%      1024  & $3019$  & $4431$ / $1.47$x & $4440$ / $1.47$x \\
%      2048  & $2480$  & $5016$ / $2.02$x & $5238$ / $2.11$x \\
%     \bottomrule
%     \end{tabular}
% }
% \end{table}

\section{Conclusion}
\label{sec:conclusion}

We present Mixer-TTS, a non-autoregressive model for speech synthesis. Both the encoder and decoder of the proposed model are based on the MLP-Mixer architecture, adapted to work with variable size input. It has pitch conditioning and a duration predictor which is trained with an unsupervised TTS alignment framework. Together with the basic model, we propose the extended version which additionally utilizes token embeddings from a pretrained LM. 

The quality of the generated speech is on par with the current state-of-the-art TTS models. The basic Mixer-TTS with HiFi-GAN vocoder achieves a MOS of 4.05, while the extended Mixer-TTS-X reaches a MOS of 4.11 (the ground truth speech has a MOS of 4.27). 
Thanks to the new design, the proposed model is fast in training and inference, which makes it an attractive candidate for speech synthesis on low-resource devices.

The model will be released as part of NeMo toolkit~\cite{kuchaiev2019nemo}.

\section{Acknowledgments}
\label{sec:acknowledgments}

The authors thank Jocelyn Huang, Jason Li, Sang-Gil Lee, Rohan Badlani, Rafael Valle and the NVIDIA AI Applications team for the helpful feedback and review. 

\bibliographystyle{IEEEbib}
\bibliography{strings,refs}

\end{document}